\documentclass[onecolumn,preprintnumbers,amsmath,amssymb,notitlepage]{revtex4}
\usepackage{graphicx}
\usepackage{epsfig}
\usepackage{amsmath}
\usepackage{amssymb}
\usepackage{dcolumn}

\begin{document}

\title{Intersection of Yang-Mills Theory with Gauge Description of General Relativity}

\author{Martin Kober${}^1$}
\email{kober@fias.uni-frankfurt.de}

\affiliation{Frankfurt Institute for Advanced Studies (FIAS),
Johann Wolfgang Goethe-Universit\"at,
Ruth-Moufang-Strasse 1, 60438 Frankfurt am Main, Germany}

\date{\today}

\begin{abstract}
An intersection of Yang-Mills theory with the gauge description of general relativity is considered.
This intersection has its origin in a generalized algebra, where the generators of the
$SO(3,1)$ group as gauge group of general relativity and the generators of a $SU(N)$ group as gauge
group of Yang-Mills theory are not separated anymore but are related by fulfilling nontrivial commutation
relations with each other. Because of the Coleman Mandula theorem this algebra cannot be postulated as
Lie algebra. As consequence, extended gauge transformations as well as an extended expression
for the field strength tensor is obtained, which contains a term consisting of products of the Yang Mills
connection and the connection of general relativity. Accordingly a new gauge invariant action
incorporating the additional term of the generalized field strength tensor is built, which depends of
course on the corresponding tensor determining the additional intersection commutation relations.
This means that the theory describes a decisively modified interaction structure between the Yang-Mills
gauge field and the gravitational field leading to a violation of the equivalence principle.
\end{abstract}


\maketitle

\section{Introduction}

One of the most important questions concerning the unification of all known interactions existing in nature is the
relation between internal symmetries referring to quantum numbers and external symmetries referring to the structure
of space-time. The reason is that these two classes of symmetries are not only a decisive criterion for the formulation
of fundamental theories, but as local symmetries they even determine the structure of the several interactions appearing
in nature, which are described as local gauge field theories with respect to these symmetries.
The interaction theories of the standard model of particle physics are based on the internal symmetries and are
formulated as special Yang-Mills theories \cite{Yang:1954ek}. In case of the electroweak theory \cite{Weinberg:1967tq} local
invariance with respect to the weak isospin is considered, whereas in quantum chromodynamics local invariance with respect to
the colour space of the quarks is considered. General relativity can be formulated as gauge theory with respect to the
Poincare group, which means that in the gauge description of gravity is considered invariance under local Lorentz
transformations or local translations
\cite{Cho:1975dh},\cite{Carmeli:1975wp},\cite{Cho:1976fr},\cite{Hayashi:1977jd},\cite{Camenzind:1978wj},\cite{Mckellar:1981hr},\cite{Grensing:1982re},\cite{Pagels:1982tc},\cite{Kawai:1985ap},\cite{Bakler:1986fu}. There also exist generalizations of
the gauge theoretic setting of gravity, for example the theories considered in
\cite{Obukhov:1987tz},\cite{Minguzzi:2001ip},\cite{Hassaine:2003vq},\cite{Krasnov:2011up},\cite{Essen:1990in},\cite{Castro:1987zk},\cite{Kober:2010sj}. This relation between the local space-time symmetries and the dynamics of general relativity means nothing else, but that the question of incorporation of gravity into a unified theory containing all interactions is directly connected to the problem of the distinguishing between internal and external symmetries.

Usually the two classes of symmetries as they appear in contemporary theoretical physics are assumed to be independent of each
other, although one should expect that in a unified theory of nature containing the interactions of the standard model as well
as gravity there has to exist some kind of relation between them, which perhaps corresponds to the relation between gravity and
the other fundamental interactions. But according to the famous theorem of Coleman and Mandula there cannot exist a Lie algebra
containing nontrivial commutation relations between the generators of internal and external symmetries \cite{Coleman:1967ad}.
Supersymmetry in a certain sense combines internal and external symmetries \cite{Wess:1973kz},\cite{Wess:1974tw}. This is
possible, because it circumvents this theorem, since its algebraic structure violates the definition of a Lie algebra,
what arises from the property that the generators of supersymmetry fulfil anticommutation relations rather than commutation
relations. Accordingly the preconditions for the Coleman Mandula theorem are not valid with respect to supersymmetry.
The ADS/CFT correspondence as duality between a supersymmetric Yang-Mills theory and a higher dimensional gravity theory
is at least a direct connection between a certain theory of gravity and a special Yang-Mills theory \cite{Maldacena:1997re}.
Other attempts to relate the two classes of symmetries, also with respect to corresponding gauge theories, are considered in
\cite{Warren:1972vk},\cite{Cho:1975bh},\cite{Huang:1976gp},\cite{Moffat:1977xk},\cite{Hennig:1981id},\cite{Sanchez:1984jm},\cite{Abe:1985xg},\cite{Lord:1987uq},\cite{Wang:1994dd},\cite{Batakis:1997wa},\cite{BottaCantcheff:2000ti},\cite{Saller:2001un}\cite{Aldaya:2006xi},\cite{Ita:2007de},\cite{Sogami:2010ws},\cite{Bern:2010yg} for example and gauge theories relating general
relativity especially to the spin degree of freedom are formulated in
\cite{Hayashi:1978xc},\cite{Dehnen:1985jj},\cite{Dehnen:1986mx},\cite{Antonowicz:1985qy}.

However, there exists another possibility to establish a relation between internal and external symmetries. This possibility
lies in the assumption of commutation relations between the generators of an internal $SU(N)$ group belonging to a certain
Yang-Mills theory and the generators of an external symmetry group like the Lorentz group, which are unequal to a linear
combination of all these generators, thus represent no Lie algebra anymore and accordingly violate the preconditions
of the Coleman Mandula theorem like supersymmetry. In this paper are assumed generalized generators of an arbitrary $SU(N)$
group and of the $SO(3,1)$ group, which fulfil commutation relations with each other being equal to a tensor in the product
space of the space of the usual $SU(N)$ generators and the space of the usual $SO(3,1)$ generators, $SU(N) \otimes SO(3,1)$.
Thus the tensor is an element of the enveloping algebra containing the usual generators of both gauge groups. The Lie algebras
of the $SO(3,1)$ group as well as the $SU(N)$ group by themselves are however assumed to remain completely unchanged and thus
the generalized algebra contains these algebras as substructure. But the additional commutation relation relating these two
gauge groups violates the structure of a Lie algebra. This means that new quantities appear defining the structure of the new
algebra and thus represent another kind of structure constants. The assumed intersection of the two symmetry groups could be
something like a first step towards a combination of these symmetries and perhaps it could be considered as a kind of
approximation to a description, where all interactions are embedded into a more general symmetry.

In accordance with this, the aim of this paper is the formulation of the corresponding intersection of Yang-Mills gauge theory
with the $SO(3,1)$ gauge description of general relativity. Therefore local gauge invariance of a matter action with respect
to the gauge group induced by the generalized algebra has to be considered. The resulting action for the matter field does not
differ from the usual one, which means that it couples in the same way to the gravitational field and the Yang-Mills field
as in the usual case, although generalized gauge transformations of the gauge fields have to be defined, since the
transformations with respect to both gauge groups influence the Yang-Mills connection as well as the connection of general
relativity. But because of the noncommutativity between the $SU(N)$ generators of the Yang-Mills theory and the $SO(3,1)$
generators one obtains a new sector in the field strength tensor containing the Yang-Mills connection as well as the connection
referring to general relativity. The corresponding interaction term within the action is assumed to be quadratic in the intersection field strength to maintain gauge invariance under the extended symmetry group containing the intersection.
The Einstein-Hilbert action as well as the usual Yang-Mills action still remain gauge invariant under the extended gauge group.
Within the generalized action appear additional interaction terms between the Yang-Mills field and the gravitational field,
which lead to a violation of the equivalence principle. The resulting theory of gravity applied to a special manifestation
of Yang-Mills theory within the standard model could yield an explanation to some phenomena in cosmology and astrophysics.

The paper is structured as follows: At the beginning the generalized algebra incorporating the additional commutation relations
containing generalized generators of the $SU(N)$ group of a Yang-Mills theory as well as the $SO(3,1)$ group is formulated.
Then the corresponding gauge theory to the generalized algebra implying generalized gauge transformations is considered.
Building of the corresponding generalized field strength tensor obtained from the covariant derivative, which is based on
connections being defined by using these new generators, leads to an intersection field strength term. This intersection field
strength combines the Yang-Mills connection and the connection of general relativity and depends on the noncommutativity
tensor defining the intersection. After this, a gauge invariant action incorporating the additional term of the field
strength tensor arising from the intersection of the gauge groups is built. Finally, the corresponding generalized energy
momentum tensor leading to a generalized Einstein field equation for the gravitational field and the generalized field
equation for the Yang-Mills field are derived.

\section{Intersection of Lorentz Group with Internal Symmetry Group}

According to the theorem of Coleman and Mandula it is not possible to construct a Lie algebra with nontrivial
commutation relations between the generators of an internal and an external symmetry group. Because of this reason
any combination of internal and external symmetries has to be expressed by an algebra, which is not a Lie algebra.
Therefore are considered commutation relations between modified generators of the $SU(N)$ group and the $SO(3,1)$
group in this paper, which are equal to a tensor in the product space of the corresponding usual generators and
thus this tensor represents itself a linear combination of generators of the $SU(N) \otimes SO(3,1)$ group.
This means that as fundamental assumption is postulated the following generalized algebra incorporating
nontrivial commutation relations of the generators of an arbitrary $SU(N)$ group as gauge group of Yang-Mills
theory, denoted with $T^A$, and the generators of the $SO(3,1)$ group as gauge group of general relativity,
denoted with $\Sigma_{ab}$:

\begin{eqnarray}
\left[T^A,T^B\right]&=&if^{ABC} T^C,\nonumber\\
\left[\Sigma_{ab},\Sigma_{cd}\right]&=&i\eta_{ac}\Sigma_{bd}
-i\eta_{bc}\Sigma_{ad}-i\eta_{ad}\Sigma_{bc}
+i\eta_{bd}\Sigma_{ac},\nonumber\\
\left[T^A,\Sigma_{ab}\right]&=&i\Lambda^{A}_{ab}=i\Gamma^{ABcd}_{ab}\tau^B \sigma_{cd}.
\label{algebra}
\end{eqnarray}
In this paper internal indices are denoted by capital latin letters, Lorentz indices are denoted by small
latin letters and space-time indices are denoted by small greek letters. 
The $f^{ABC}$ describe the structure constants of the $SU(N)$ group and $\eta_{ab}$ denotes the Minkowski metric. Besides
there has been introduced the noncommutativity parameter $\Lambda_{ab}^{A}$ as a linear combination of the generators
of $SU(N) \otimes SO(3,1)$ defined by the coefficients $\Gamma^{ABcd}_{ab}$.
The quantities $\tau^{A}$ and $\sigma_{ab}$ are the usual generators of the $SU(N)$ and the $SO(3,1)$ group respectively
and thus fulfil the usual commutation relations without intersection,

\begin{eqnarray}
\left[\tau^A,\tau^B\right]&=&if^{ABC} \tau^C,\nonumber\\
\left[\sigma_{ab},\sigma_{cd}\right]&=&i\eta_{ac}\sigma_{bd}
-i\eta_{bc}\sigma_{ad}-i\eta_{ad}\sigma_{bc}
+i\eta_{bd}\sigma_{ac},\nonumber\\
\left[\tau^A,\sigma_{ab}\right]&=&0.
\label{usual_algebra}
\end{eqnarray}
The generalized generators $T^A$ and $\Sigma_{ab}$ obeying the generalized algebra ($\ref{algebra}$) can be represented
as linear combination of the usual generators $\tau^A$ and $\sigma_{ab}$ defined in ($\ref{usual_algebra}$),

\begin{equation}
T^A=\chi^{AB}\tau^B,\quad \Sigma_{ab}=M_{ab}^{\ \ cd}\sigma_{cd},
\label{definition_generators}
\end{equation}
where the coefficients $\chi^{AB}$ belonging to the usual generators of the $SU(N)$ group as well as the coefficients
$M_{ab}^{\ \ cd}$ belonging to the usual generators of the $SO(3,1)$ group are not usual numbers, but fulfil nontrivial
commutation relations with each other,

\begin{eqnarray}
\left[\chi^{AB},M_{ab}^{\ \ cd}\right]=i\Gamma^{ABcd}_{ab},\quad
\left[\chi^{AB},\chi^{CD}\right]=0,\quad
\left[M_{ab}^{\ \ cd},M^{\ \ ef}_{gh}\right]=0.
\label{algebra_coefficients}
\end{eqnarray}
Since the components of $\chi^{AB}$ as well as the ones of $M_{ab}^{\ \ cd}$ are numbers but fulfil nontrivial commutation relations anyhow, they are quantities similar to grassmann numbers. The algebra ($\ref{algebra_coefficients}$)
of course implies the canonical commutation relations between the generalized generators defined in ($\ref{algebra}$),

\begin{eqnarray}
\left[T^A,\Sigma_{ab}\right]=\left[\chi^{AB}\tau^B,M^{\ \ cd}_{ab}\sigma_{cd}\right]
=\left[\chi^{AB},M^{\ \ cd}_{ab}\right]\tau^B \sigma_{cd}
=i\Gamma^{ABcd}_{ab}\tau^B \sigma_{cd}= i\Lambda^A_{ab}.
\end{eqnarray}
Since the generalized generators $\Sigma_{ab}$ as well as the usual generators $\sigma_{ab}$ are antisymmetric
with respect to the indices, the tensor $M_{ab}^{\ \ cd}$ belonging to the generators of the $SO(3,1)$ group is
also antisymmetric with respect to the first pair and the second pair of indices:

\begin{equation}
M_{ab}^{\ \ cd}=-M_{ba}^{\ \ cd}=-M_{ab}^{\ \ dc}=M_{ba}^{\ \ dc}.
\label{antisymmetry}
\end{equation} 
The tensors $\chi^{AB}$ and $M_{ab}^{\ \ cd}$ are further assumed to obey the following relations:

\begin{equation}
\chi^{AB}=\left(\chi^{-1}\right)^{BA},\quad M_{ab}^{\ \ cd}=\left(M^{-1}\right)_{\ \ ab}^{cd},
\label{symmetry_coefficients}
\end{equation}
which lead to certain trace properties for the new generators defined by the generalized algebra ($\ref{algebra}$), which
means that the generalized $SU(N)$ generators fulfil the same and the $SO(3,1)$ fulfil similar trace properties as the
usual generators ($\ref{usual_algebra}$) as will be shown below. That the generators $\Sigma_{ab}$ with each other
still fulfil the commutation relations of the $SO(3,1)$ group and the generators $T^A$ with each other still fulfil
the commutation relations of the $SU(N)$ group implies that the following relations for the coefficients have to be
valid, too:

\begin{equation}
f^{IJK}\tau^K=\chi^{AI}\chi^{BJ}\chi^{CK}f^{ABC}\tau^K,\quad
\eta_{ik}\sigma_{jl}=M^{ab}_{\ \ ij} M^{cd}_{\ \ kl} M_{bd}^{\ \ fh}
\eta_{ac}\sigma_{fh}.
\label{conditions_coefficients}
\end{equation}
By using ($\ref{algebra}$),($\ref{usual_algebra}$),($\ref{definition_generators}$),($\ref{algebra_coefficients}$)
and ($\ref{symmetry_coefficients}$), the conditions ($\ref{conditions_coefficients}$) on the coefficients can be
derived as follows:

\begin{eqnarray}
&&\left[T^A,T^B\right]=if^{ABC}T^C\quad\Leftrightarrow\quad
\left[\chi^{AD}\tau^D,\chi^{BE}\tau^E\right]=if^{ABC}\chi^{CG}\tau^G
\quad\Leftrightarrow\quad\chi^{AD}\chi^{BE}\left[\tau^D,\tau^E\right]=i\chi^{CG}f^{ABC}\tau^G\nonumber\\
&&\Leftrightarrow\quad i\chi^{AD}\chi^{BE}f^{DEH}\tau^H=i\chi^{CG}f^{ABC}\tau^G\quad
\Leftrightarrow\quad i\left(\chi^{-1}\right)^{IA}\left(\chi^{-1}\right)^{JB}\chi^{AD}\chi^{BE}f^{DEH}\tau^H
=i\left(\chi^{-1}\right)^{IA}\left(\chi^{-1}\right)^{JB}\chi^{CG}f^{ABC}\tau^G\nonumber\\
&&\Leftrightarrow\quad i\delta^{ID}\delta^{JE}f^{DEH}\tau^H
=i\left(\chi^{-1}\right)^{IA}\left(\chi^{-1}\right)^{JB}\chi^{CG}f^{ABC}\tau^G
\quad\Leftrightarrow\quad if^{IJK}\tau^K
=i\chi^{AI}\chi^{BJ}\chi^{CK}f^{ABC}\tau^K,
\end{eqnarray}
and the second relation can be derived analogously:

\begin{eqnarray}
&&\left[\Sigma_{ab},\Sigma_{cd}\right]
=i\eta_{ac}\Sigma_{bd}-i\eta_{bc}\Sigma_{ad}
-i\eta_{ad}\Sigma_{bc}+i\eta_{bd}\Sigma_{ac}
\nonumber\\
&&\Leftrightarrow\quad\left[M_{ab}^{\ \ ef}\sigma_{ef},
M_{cd}^{\ \ gh}\sigma_{gh}\right]
=i\eta_{ac}M_{bd}^{\ \ fh}\sigma_{fh}
-i\eta_{bc}M_{ad}^{\ \ eh}\sigma_{eh}
-i\eta_{ad}M_{bc}^{\ \ fg}\sigma_{fg}
+i\eta_{bd}M_{ac}^{\ \ eg}\sigma_{eg}
\nonumber\\
&&\Leftrightarrow\quad M_{ab}^{\ \ ef}M_{cd}^{\ \ gh}
\left[\sigma_{ef},\sigma_{gh}\right]
=i\eta_{ac}M_{bd}^{\ \ fh}\sigma_{fh}
-i\eta_{bc}M_{ad}^{\ \ eh}\sigma_{eh}
-i\eta_{ad}M_{bc}^{\ \ fg}\sigma_{fg}
+i\eta_{bd}M_{ac}^{\ \ eg}\sigma_{eg}
\nonumber\\
&&\Leftrightarrow\quad M_{ab}^{\ \ ef}M_{cd}^{\ \ gh}
\left(i\eta_{eg}\sigma_{fh}-i\eta_{fg}\sigma_{eh}
-i\eta_{eh}\sigma_{fg}+i\eta_{fh}\sigma_{eg}\right)
\nonumber\\
&&\quad\quad\quad=i\eta_{ac}M_{bd}^{\ \ fh}\sigma_{fh}
-i\eta_{bc}M_{ad}^{\ \ eh}\sigma_{eh}
-i\eta_{ad}M_{bc}^{\ \ fg}\sigma_{fg}
+i\eta_{bd}M_{ac}^{\ \ eg}\sigma_{eg}
\nonumber\\
&&\Leftrightarrow\quad iM_{ab}^{\ \ ef}M_{cd}^{\ \ gh}\eta_{eg}\sigma_{fh}
=i\eta_{ac}M_{bd}^{\ \ fh}\sigma_{fh}
\nonumber\\
&&\Leftrightarrow\quad
i\left(M^{-1}\right)_{ij}^{\ \ ab}\left(M^{-1}\right)_{kl}^{\ \ cd}
M_{ab}^{\ \ ef}M_{cd}^{\ \ gh}\eta_{eg}\sigma_{fh}
=i\left(M^{-1}\right)_{ij}^{\ \ ab}\left(M^{-1}\right)_{kl}^{\ \ cd}
\eta_{ac}M_{bd}^{\ \ fh}\sigma_{fh}
\nonumber\\
&&\Leftrightarrow\quad
i\delta^e_i \delta^f_j \delta^g_k \delta^h_l \eta_{eg}\sigma_{fh}
=i\left(M^{-1}\right)_{ij}^{\ \ ab}\left(M^{-1}\right)_{kl}^{\ \ cd}
\eta_{ac} M_{bd}^{\ \ fh}\sigma_{fh}
\nonumber\\
&&\Leftrightarrow\quad
i\eta_{ik}\sigma_{jl}
=i\left(M^{-1}\right)_{ij}^{\ \ ab}\left(M^{-1}\right)_{kl}^{\ \ cd}
\eta_{ac}M_{bd}^{\ \ fh}\sigma_{fh}
\nonumber\\
&&\Leftrightarrow\quad
i\eta_{ik}\sigma_{jl}
=iM^{ab}_{\ \ ij}M^{cd}_{\ \ kl}M_{bd}^{\ \ fh}\eta_{ac}\sigma_{fh}.
\end{eqnarray}
In the next sections will be developed the corresponding gauge theory to the gauge group based on the algebra
considered in this section ($\ref{algebra}$). To maintain gauge invariance of the corresponding action, which
will be built, it is important to calculate the trace of the product of two of the generalized generators,
${\rm tr}\left[T^A T^B\right]$ and ${\rm tr}\left[\Sigma_{ab}\Sigma_{cd}\right]$ respectively, from the
trace of the product of two of the usual generators, which is given by

\begin{equation}
{\rm tr}\left[\tau^A \tau^B\right]=\frac{1}{2}\delta^{AB}\quad,\quad
{\rm tr}\left[\sigma_{ab}\sigma_{cd}\right]=\eta_{ac}\eta_{bd}-\eta_{ad}\eta_{bc}.
\label{trace_usual_generators}
\end{equation}
By using ($\ref{symmetry_coefficients}$) and ($\ref{trace_usual_generators}$) this can be done for
the generalized generators of the $SU(N)$ group as follows :

\begin{equation}
{\rm tr}\left[T^A T^B\right]={\rm tr}\left[\chi^{AC}\tau^C\chi^{BD}\tau^D\right]
=\chi^{AC}\chi^{BD}{\rm tr}\left[\tau^C\tau^D\right]=\frac{1}{2}\chi^{AC}\chi^{BD}\delta^{CD}
=\frac{1}{2}\chi^{AC}\chi^{BC}=\frac{1}{2}\chi^{AC}\left(\chi^{-1}\right)^{CB}
=\frac{1}{2}\delta^{AB},
\label{trace_generalized_generators_Yang-Mills}
\end{equation}
and for the trace of the generalized generators of the $SO(3,1)$ group it can be done analogously:

\begin{eqnarray}
{\rm tr}\left[\Sigma_{ab}\Sigma_{cd}\right]&=&
{\rm tr}\left[M^{\ \ ef}_{ab}\sigma_{ef}M^{\ \ kl}_{cd}\sigma_{kl}\right]
=M^{\ \ ef}_{ab}M^{\ \ kl}_{cd}
{\rm tr}\left[\sigma_{ef}\sigma_{kl}\right]
=M^{\ \ ef}_{ab}M^{\ \ kl}_{cd}
\left(\eta_{ek}\eta_{fl}-\eta_{el}\eta_{fk}\right)\nonumber\\
&=&M_{ab}^{\ \ ef} M_{cdef}
-M_{ab}^{\ \ ef} M_{cdfe}
=2M_{ab}^{\ \ ef} M_{cdef}
=2M_{ab}^{\ \ ef} \left(M^{-1}\right)_{efcd}\nonumber\\
&=&2M_{ab}^{\ \ ef} \left(M^{-1}\right)_{ef}^{\ \ kl}\eta_{kc}\eta_{ld}
=2\delta_a^k \delta_b^l \eta_{kc}\eta_{ld}
=2\eta_{ac}\eta_{bd}.
\label{trace_generalized_generators_Lorentz}
\end{eqnarray}
%where has been used the antisymmetry property of $M_{ab}^{cd}$ given in ($\ref{antisymmetry}$),
%which also implies that $M_{ab\ c}^{\ \ c}$=0.

\section{Corresponding Generalization of Gauge Transformations}

In the last section has been introduced a generalized algebra for the generators of a $SU(N)$ group and the $SO(3,1)$ group
leading to an intersection between these two groups. In this section the corresponding gauge theory based on the combined gauge
group belonging to this generalized algebra is considered. Since local gauge invariance is postulated with respect to a
fermionic matter field $\psi$, which dynamics is described by the Dirac equation, the $SO(3,1)$ group has to be represented
in the Dirac spinor space implying that the usual generators $\sigma_{ab}$ take the following shape:
$\sigma_{ab}=-\frac{i}{4}\left[\gamma_a,\gamma_b\right]$.
This implies for the generalized generators $\Sigma_{ab}$ related to the usual ones by ($\ref{definition_generators}$)
$\Sigma_{ab}=-\frac{i}{4}M_{ab}^{\ \ cd}\left[\gamma_c,\gamma_d\right]$. The matter field $\psi$
has to be assumed to have another internal degree of freedom a $SU(N)$ group refers to and accordingly the corresponding
Dirac equation is invariant under global Yang-Mills gauge transformations containing the generalized generators,
$\mathcal{U}_{YM}=\exp\left(i\alpha^A T^A \right)$, as well as global Lorentz gauge transformations containing
the generalized generators, $\mathcal{U}_{L}=\exp\left(i\varphi^{ab}\Sigma_{ab}\right)$.
Accordingly it is also invariant under a combined gauge transformation, which has to symmetrized because
of the noncommutativity of the operators having its origin in the presupposed noncommutativity of the
generators of the two gauge groups, $\left[T^A,\Sigma_{ab}\right] \neq 0$,

\begin{eqnarray}
\mathcal{U}_G&=&\frac{\mathcal{U}_{YM}\mathcal{U}_L+\mathcal{U}_L\mathcal{U}_{YM}}{2}
=\frac{\exp\left(i\alpha^A T^A\right)\exp\left(i\varphi^{ab}\Sigma_{ab}\right)
+\exp\left(i\varphi^{ab}\Sigma_{ab}\right)\exp\left(i\alpha^A T^A\right)}{2}\nonumber\\&&
=\frac{\exp\left(i\alpha^A T^A+i\varphi^{ab}\Sigma_{ab}-\frac{i}{2}\alpha^A\varphi^{ab}\Lambda^{A}_{ab}+...\right)
+\exp\left(i\varphi^{ab}\Sigma_{ab}+i\alpha^A T^A+\frac{i}{2}\alpha^A\varphi^{ab}\Lambda^{A}_{ab}+...\right)}{2}\nonumber\\&&
={\bf 1}+i\alpha^A T^A+i\varphi^{ab}\Sigma_{ab}+\mathcal{O}\left(\alpha^2,\varphi^2\right),
\label{transformation_operator}
\end{eqnarray}
where has to be used the Baker-Campbell-Hausdorff formula, which reads as following:

\begin{equation}
\exp(X)\exp(Y)=\exp(X+Y+\frac{1}{2}[X,Y]+\frac{1}{12}[X,[X,Y]]-\frac{1}{12}[Y,[X,Y]]+...).
\end{equation}
Except that the generators fulfil the more general algebra, the infinitesimal element of the transformation
operator ($\ref{transformation_operator}$) corresponds to the usual transformation operator, because to 
the first order in $\alpha$ and $\varphi$ the additional expressions of the two terms with different
order of the separated transformation operators cancel out. 
Since $\bar \psi=\psi^{\dagger}\gamma^0$ is the adjoint spinor to $\psi$ in the Dirac spinor space,
$\gamma^0 \mathcal{U}^{\dagger}_L\gamma^0$ is the adjoint operator to $\mathcal{U}_L$, at least in the
infinitesimal case considered throughout this paper, in which it is unitary meaning that $\gamma^0 \mathcal{U}^{\dagger}_L\gamma^0 \mathcal{U}_L={\bf 1}$ and accordingly $\gamma^0 \mathcal{U}^{\dagger}_G\gamma^0 \mathcal{U}_G={\bf 1}$, which means $\mathcal{U}_G^{-1}=\gamma^0\mathcal{U}_G^{\dagger}\gamma^0$.
To maintain local gauge invariance with respect to the combined gauge group correspopnding to the
transformation operator ($\ref{transformation_operator}$), one has to consider the following
Lagrangian of a fermionic matter field:

\begin{equation}
\mathcal{L}_M=e\bar \psi\left(i\gamma^m \mathcal{D}_m-m\right)\psi,
\label{gauge_invariant_matter_Lagrangian}
\end{equation}
where has been defined $e=\det\left[e_\mu^m \right]$ and the covariant derivative $\mathcal{D}_m$ corresponds
to the usual covariant derivative with respect to an internal $SU(N)$ symmetry and the external $SO(3,1)$ symmetry
with the usual generators obeying ($\ref{usual_algebra}$) replaced by the generalized generators
obeying ($\ref{algebra}$),

\begin{equation}
\mathcal{D}_m=e^\mu_m\left(\partial_\mu+iA_\mu^A T^A+\frac{i}{2}\omega_\mu^{ab}\Sigma_{ab}\right)
=e^\mu_m\left(\partial_\mu+iA_\mu^A\chi^{AB}\tau^B+\frac{i}{2}\omega_\mu^{ab}M^{cd}_{ab}\sigma_{cd}\right).
\label{covariant_derivative}
\end{equation}
The coefficients of the spin connection $\omega_\mu^{ab}$ are related to the tetrad field $e_m^\mu$ according to 

\begin{equation}
\omega_\mu^{ab}=2e^{\nu a}\partial_\mu e_\nu^b-2e^{\nu b} \partial_\mu e_\nu^a
-2e^{\nu a}\partial_\nu e_\mu^b+2e^{\nu b} \partial_\nu e_\mu^a
+e_{\mu c} e^{\nu a}e^{\sigma b}\partial_\sigma e_\nu^c
-e_{\mu c} e^{\nu a}e^{\sigma b}\partial_\nu e_\sigma^c.
\label{connection_tetrad}
\end{equation}
Accordingly the Lagrangian ($\ref{gauge_invariant_matter_Lagrangian}$) is invariant under the following
local symmetry transformation containing the transformation operator defined in ($\ref{transformation_operator}$)
with a space-time dependent gauge parameter:

\begin{eqnarray}
\psi \longrightarrow \mathcal{U}_G \psi,\quad \mathcal{D}_m \longrightarrow
\phi^{\ n}_m \mathcal{U}_G \mathcal{D}_n \gamma^0\mathcal{U}_G^{\dagger}\gamma^0
,\quad e^\mu_m \longrightarrow \phi^{\ n}_m e^\mu_n,\quad
\gamma^m \longrightarrow \phi^m_{\ n} \mathcal{U}_G \gamma^n \gamma^0 \mathcal{U}_G^{\dagger}\gamma^0,
\label{transformations}
\end{eqnarray}
where $\phi^m_{\ n}$ denotes a non infinitesimal Lorentz transformation matrix describing the transformation
of Lorentz indices: $x^m \longrightarrow \phi^m_{\ n} x^n$. This means that the gauge fields transform as

\begin{eqnarray}
A_\mu \longrightarrow \mathcal{U}_G A_\mu \gamma^0 \mathcal{U}_G^{\dagger}\gamma^0-\gamma^0 \mathcal{U}_G^{\dagger}\gamma^0\partial_\mu \alpha\ \mathcal{U}_G,\quad
\omega_\mu \longrightarrow \mathcal{U}_G \omega_\mu \gamma^0 \mathcal{U}_G^{\dagger}\gamma^0-\gamma^0 \mathcal{U}_G^{\dagger}\gamma^0\partial_\mu \omega\ \mathcal{U}_G,\quad
e_m^\mu \longrightarrow \phi_m^{\ n} e_n^\mu,
\end{eqnarray}
if infinitesimal transformations are considered.
The transformation rules ($\ref{transformations}$) imply that the matter field and the gauge potentials
appearing in the covariant derivative ($\ref{covariant_derivative}$) have to transform according to

\begin{eqnarray}
\psi &\longrightarrow& \left({\bf 1}+i\alpha+i\varphi\right)\psi
+\mathcal{O}\left(\alpha^2,\varphi^2\right),\nonumber\\
A_\mu &\longrightarrow& A_\mu-\partial_\mu \alpha+i[\alpha,A_\mu]+i[\varphi,A_\mu]
+\mathcal{O}\left(\alpha^2,\varphi^2\right),\nonumber\\
\omega_\mu &\longrightarrow& \omega_\mu-\partial_\mu \varphi+i[\varphi,\omega_\mu]+i[\alpha,\omega_\mu]
+\mathcal{O}\left(\alpha^2,\varphi^2\right),\nonumber\\
e_m^\mu &\longrightarrow& e_m^\mu+\varphi_m^{\ n} e_n^\mu+\mathcal{O}\left(\varphi^2\right),
\end{eqnarray}
with $\alpha=\alpha^A T^A$ and $\varphi=\varphi^{ab}\Sigma_{ab}$.
The invariance of the Dirac Lagrangian ($\ref{gauge_invariant_matter_Lagrangian}$) containing the 
generalized covariant derivative ($\ref{covariant_derivative}$) under local gauge transformations
of the shape ($\ref{transformations}$) can in analogy to the usual case be seen as follows:

\begin{equation}
e\bar \psi\left(i\gamma^\mu \mathcal{D}_\mu-m\right)\psi
\longrightarrow e\bar \psi \gamma^0 \mathcal{U}_G^{\dagger}\gamma^0\left(i\phi^m_{\ n}\mathcal{U}_G\gamma^n \gamma^0 \mathcal{U}_G^{\dagger} \gamma^0 \phi_m^{\ p}\mathcal{U}_G\mathcal{D}_p \gamma^0 \mathcal{U}_G^{\dagger} \gamma^0
-m\right)\mathcal{U}_G\psi
=e\bar \psi\left(i\gamma^\mu \mathcal{D}_\mu-m\right)\psi,
\end{equation}
since $\gamma^0 \mathcal{U}_G^{\dagger} \gamma^0 \mathcal{U}_G=\mathcal{U}_G^{-1}\mathcal{U}_G={\bf 1}$
%, what is only valid for $\mathcal{U}_G$ but not for the product $\mathcal{U}_{YM}\otimes \mathcal{U}_L$,
and $\phi^m_{\ n} \phi^{\ p}_m=\delta^p_n$.
%It is important to mention that it is necessary that the gauge transformation of the gamma matrices according to
%($\ref{transformations}$) is performed by the complete transformation operator $\mathcal{U}_G$ containing the Lorentz
%transformation rather than just with the separated Lorentz transformation operator $\mathcal{U}_L$. This holds
%because for single gamma-matrices both transformations are equivalent, since the Yang-Mills gauge transformation
%within $\mathcal{U}_G$ has no effect on the gamma matrices and cancels out. With respect to a gauge transformation
%of the complete Dirac Lagrangian by the combined gauge group the appearance of separated Lorentz transformations
%operators $\mathcal{U}_L$ would spoil gauge invariance for non infinitesimal gauge transformations because of
%$\left[T^A,\Sigma_{ab}\right] \neq 0$.

\section{Generalized Dynamics Incorporating the Intersection Field Strength}

In the last section has been shown that the dynamics of fermionic matter fields is not modified under incorporation
of the generalized combined gauge group, although the gauge transformation conditions have to be extended. But this holds not for the dynamics of the gauge fields themselves. The reason is that the algebraic properties of the covariant derivative
are changed and this leads to an additional term of the field strength tensor, from which the action of any gauge field is
built usually. This means that a generalized Lagrangian describing generalized dynamics of the Yang-Mills field and the
gravitational field has to be constructed by incorporating the additional term induced by the generalized covariant
derivative ($\ref{covariant_derivative}$). Therefore the generalized field strength tensor has to be calculated first,
which is defined as usual as the commutator of the covariant derivatives and accordingly reads as follows:

\begin{eqnarray}
\mathcal{F}_{mn}&=&-i[\mathcal{D}_m,\mathcal{D}_n]\nonumber\\
&=&-i[e_m^\mu \mathcal{D}_\mu, e_n^\nu \mathcal{D}_\nu]\nonumber\\
&=&-i\left[e_m^\mu\left(\mathcal{D}_\mu e_n^\nu\right)\mathcal{D}_\nu-e_n^\nu\left(\mathcal{D}_\nu e_m^\mu\right)\mathcal{D}_\mu
+e^\mu_m e^\nu_n \mathcal{D}_\mu \mathcal{D}_\nu-e^\mu_m e^\nu_n \mathcal{D}_\nu \mathcal{D}_\mu\right]\nonumber\\
&=&-ie^\mu_m e^\nu_n T_{\mu\nu}^{\rho}\mathcal{D}_\rho+e_m^\mu e_n^\nu \frac{1}{2}R_{\mu\nu}^{ab}\Sigma_{ab}
+e_m^\mu e_n^\nu G_{\mu\nu}^A T^A-e_m^\mu e_n^\nu H_{\mu\nu}^{Aab}\Lambda_{ab}^{A},
\end{eqnarray}
where the following expressions for the several field strength sectors have been defined: 

\begin{eqnarray}
T_{\mu\nu}^{\rho}&=&\mathcal{D}_\mu e^\rho_\nu-\mathcal{D}_\nu e^\rho_\mu,\nonumber\\
R_{\mu\nu}^{ab}&=&\partial_\mu \omega_\nu^{ab}-\partial_\nu \omega_\mu^{ab}
+\omega_\mu^{ac}\omega_\nu^{cb}-\omega_\nu^{ac}\omega_\mu^{cb},\nonumber\\
G_{\mu\nu}^A&=&\partial_\mu A_\nu^A-\partial_\nu A_\mu^A-f^{ABC}A_\mu^B A_\nu^C,\nonumber\\
H_{\mu\nu}^{Aab}&=&A_\mu^A \omega_\nu^{ab}-A_\nu^A \omega_\mu^{ab}.
\label{sectors_field_strength_tensor}
\end{eqnarray}
From the noncommutativity of the generators of the Yang-Mills gauge group with the generators of the
Lorentz group the new field strength tensor $H_{\mu\nu}^{Aab}$ arises besides the torsion $T_{\mu\nu}^{\rho}$,
which is assumed to vanish, $T_{\mu\nu}^{\rho}=0$, the Riemann tensor $R_{\mu\nu}^{ab}$ and the Yang-Mills
field strength tensor  $G_{\mu\nu}^A$. Because of the transformation property of the covariant derivative
($\ref{transformations}$) the field strength transforms analogously as follows:

\begin{eqnarray}
\mathcal{F}_{mn}=-i\left[\mathcal{D}_m,\mathcal{D}_n\right]
&\longrightarrow&
-i\left[\mathcal{U}_G \mathcal{D}_m \gamma^0 \mathcal{U}_G^{\dagger}\gamma^0,\mathcal{U}_G \mathcal{D}_n \gamma^0 \mathcal{U}_G^{\dagger} \gamma^0\right] 
=-i\mathcal{U}_G\left[\mathcal{D}_m,\mathcal{D}_n\right]\gamma^0 \mathcal{U}_G^{\dagger} \gamma^0
=\mathcal{U}_G \mathcal{F}_{mn} \gamma^0 \mathcal{U}_G^{\dagger} \gamma^0.
\label{gauge_transformation_field_strength}
\end{eqnarray}
The transformation rule ($\ref{gauge_transformation_field_strength}$) of course implies that all parts of the field
strength tensor including the additional term $e_m^\mu e_n^\nu H_{\mu\nu}^{Aab}\Lambda_{ab}^{A}$
transform completely analogously to each other, which means

\begin{eqnarray}
e^\mu_m e^\nu_n T_{\mu\nu}^{\rho} \mathcal{D}_\rho &\longrightarrow&
e^\mu_m e^\nu_n \mathcal{U}_G \left(T_{\mu\nu}^{\rho} \mathcal{D}_\rho\right) \gamma^0 \mathcal{U}_G^{\dagger} \gamma^0,
\nonumber\\
e_m^\mu e_n^\nu \frac{1}{2}R_{\mu\nu}^{ab}\Sigma_{ab}
&\longrightarrow& e_m^\mu e_n^\nu \mathcal{U}_G \left(\frac{1}{2}R_{\mu\nu}^{ab}\Sigma_{ab}\right) \gamma^0 \mathcal{U}_G^{\dagger} \gamma^0,
\nonumber\\
e^\mu_m e^\nu_n G_{\mu\nu}^A T^A &\longrightarrow& e^\mu_m e^\nu_n \mathcal{U}_G \left(G_{\mu\nu}^A T^A\right) \gamma^0 \mathcal{U}_G^{\dagger} \gamma^0,
\nonumber\\
e_m^\mu e_n^\nu H_{\mu\nu}^{Aab}\Lambda_{ab}^{A} &\longrightarrow&
e_m^\mu e_n^\nu \mathcal{U}_G \left(H_{\mu\nu}^{Aab}\Lambda_{ab}^{A}\right) \gamma^0 \mathcal{U}_G^{\dagger} \gamma^0.
\label{transformation_field_strength_sectors}
\end{eqnarray}
To construct a Lagrangian containing the intersection field strength, the quantity $\mathcal{H}^{Aab}_{\mu\nu}$
incorporating the noncommutativity parameter of the generalized algebra for the generators ($\ref{algebra}$)
is introduced, which is defined according to

\begin{equation}
H_{\mu\nu}^{Aab}\Lambda^{A}_{ab}
=H_{\mu\nu}^{Aab}\Gamma^{ABcd}_{ab}\tau^{B}\sigma_{cd}
\equiv \mathcal{H}^{Bcd}_{\mu\nu}\tau^{B}\sigma_{cd}.
\label{reformulation_intersection_field_strength}
\end{equation}
The generalized action for the gauge fields $\mathcal{S}_G$ based on the several sectors of the field strength tensor ($\ref{sectors_field_strength_tensor}$), which is postulated in this paper, consists of the usual Einstein-Hilbert action,
the usual action for a Yang-Mills field on curved space-time and an intersection action containing the new sector of
the field strength tensor defined in ($\ref{sectors_field_strength_tensor}$) and ($\ref{reformulation_intersection_field_strength}$) respectively: $\mathcal{S}_{G}=S_{EH}+S_{YM}+S_{Int}$,
where $\mathcal{S}_{EH}$ denotes the Einstein-Hilbert action, $\mathcal{S}_{YM}$ denotes the Yang-Mills action
and $\mathcal{S}_{Int}$ denotes the intersection action. The intersection action is assumed to be quadratic and
formulated in analogy to the Yang-Mills action and thus the complete action for the gauge fields reads

\begin{equation}
\mathcal{S}_G=\int d^4 x\ e\left(\frac{1}{16 \pi G}e^\mu_a e^\nu_b R_{\mu\nu}^{ab}
+\frac{1}{4}e^{\mu a} e^\rho_a e^{\nu b} e^\sigma_b G_{\mu\nu}^A G_{\rho\sigma}^A
+\frac{1}{4} e^{\mu a}e^{\rho}_{a}e^{\nu b}e^{\sigma}_{b}\mathcal{H}_{\mu\nu}^{Acd}
\mathcal{H}^{A}_{\rho\sigma cd}\right).
\label{action_gauge_fields}
\end{equation}
The gauge coupling constant of Yang-Mills theory is assumed to be equal to one in this paper
in contrast to the gravitational constant $G$.
To justify the action for the gauge fields ($\ref{action_gauge_fields}$), it has to be shown that this postulated
action including the intersection term is gauge invariant with respect to the gauge transformations ($\ref{transformation_field_strength_sectors}$). Since the generators have extended properties, gauge
invariance has not only to be shown with respect to the intersection Lagrangian, but also with respect to the
Yang-Mills Lagrangian and the Einstein-Hilbert Lagrangian. To perform the gauge transformations, the
several terms belonging to the action have to be rewritten by incorporating the generators and building
the traces. Thus the trace properties of the products of two of the generators considered in
($\ref{trace_usual_generators}$),($\ref{trace_generalized_generators_Yang-Mills}$) and ($\ref{trace_generalized_generators_Lorentz}$) become important here. At the beginning is
considered gauge invariance of the Einstein-Hilbert term. The following calculation shows
that the Einstein-Hilbert term remains the same after a gauge transformation:

\begin{eqnarray}
&&e^\mu_a e^\nu_b R_{\mu\nu}^{ab}=e^{\mu c}e^{\nu d}R_{\mu\nu}^{ab}
\eta_{ac}\eta_{bd}=\frac{1}{2}{\rm tr}\left[e^{\mu c} e^{\nu d} 
\Sigma_{cd} R_{\mu\nu}^{ab}\Sigma_{ab}\right]\nonumber\\
&&\longrightarrow
\frac{1}{2}{\rm tr}\left[\mathcal{U}_G \left(e^{\mu c} e^{\nu d} \Sigma_{cd}\right)\gamma^0 \mathcal{U}^{\dagger}_G \gamma^0
\mathcal{U}_G \left(R_{\mu\nu}^{ab}\Sigma_{ab}\right)\gamma^0 \mathcal{U}^{\dagger}_G \gamma^0\right]
=\frac{1}{2} e^{\mu c} e^{\nu d} R_{\mu\nu}^{ab}\ {\rm tr}\left[\mathcal{U}_G \Sigma_{ab}
\Sigma_{cd}\gamma^0 \mathcal{U}^{\dagger}_G \gamma^0\right]\nonumber\\
&&\quad\quad=\frac{1}{2} e^{\mu c} e^{\nu d} R_{\mu\nu}^{ab}\ {\rm tr}\left[\left({\bf 1}+i\alpha^A T^A+i\varphi^{ef}\Sigma_{ef}\right)\Sigma_{ab}\Sigma_{cd}
\left({\bf 1}-i\alpha^A T^A-i\varphi^{ef}\Sigma_{ef}\right)\right]
+\mathcal{O}\left(\alpha^2,\varphi^2\right)\nonumber\\
&&\quad\quad=\frac{1}{2} e^{\mu c} e^{\nu d} R_{\mu\nu}^{ab}\ {\rm tr}\left[\Sigma_{ab}\Sigma_{cd}
+i\alpha^A\left(T^A \Sigma_{ab}\Sigma_{cd}-\Sigma_{ab}\Sigma_{cd}T^A\right)
+i\varphi^{ef}\left(\Sigma_{ef}\Sigma_{ab}\Sigma_{cd}
-\Sigma_{ab}\Sigma_{cd}\Sigma_{ef}\right)\right]\nonumber\\
&&\quad\quad=\frac{1}{2} e^{\mu c} e^{\nu d} R_{\mu\nu}^{ab}\ {\rm tr}\left[\Sigma_{ab}\Sigma_{cd}
+i\alpha^A\left(T^A \Sigma_{ab}\Sigma_{cd}-\Sigma_{ab}\Sigma_{cd}T^A\right)
+i\varphi^{ef}\left(i\eta_{ea}\Sigma_{fb}-i\eta_{fa}\Sigma_{eb}
-i\eta_{eb}\Sigma_{fa}+i\eta_{fb}\Sigma_{ea}\right)\Sigma_{cd}
\right.\nonumber\\&&\left.\quad\quad\quad\quad\quad\quad\quad\quad\quad\quad
-i\varphi^{ef}\Sigma_{ab}\left(i\eta_{ce}\Sigma_{df}
-i\eta_{de}\Sigma_{cf}-i\eta_{cf}\Sigma_{de}
+i\eta_{df}\Sigma_{ce}\right)\right]\nonumber\\
&&\quad\quad=\frac{1}{2} e^{\mu c} e^{\nu d} R_{\mu\nu}^{ab}\ {\rm tr}\left[\Sigma_{ab}\Sigma_{cd}
+i\alpha^A\left(T^A \Sigma_{ab}\Sigma_{cd}-\Sigma_{ab}\Sigma_{cd}T^A\right)
+i\varphi^{ef}\left(2i\eta_{ea}\Sigma_{fb}
-2i\eta_{eb}\Sigma_{fa}\right)\Sigma_{cd}
\right.\nonumber\\&&\left.\quad\quad\quad\quad\quad\quad\quad\quad\quad\quad
-i\varphi^{ef}\Sigma_{ab}\left(2i\eta_{ce}\Sigma_{df}
-2i\eta_{de}\Sigma_{cf}\right)\right]\nonumber\\
&&\quad\quad=e^{\mu c} e^{\nu d} R_{\mu\nu}^{ab}
\left[\eta_{ac} \eta_{bd}
+i\alpha^A\left(T^A \eta_{ac} \eta_{bd}
-\eta_{ac} \eta_{bd} T^A\right)
+i\varphi^{ef}\left(
2i\eta_{ea}\eta_{fc}\eta_{bd}
-2i\eta_{eb}\eta_{fc}\eta_{ad}\right)
\right.\nonumber\\&&\left.\quad\quad\quad\quad\quad\quad\quad\quad
-i\varphi^{ef}
\left(2i\eta_{ce}\eta_{ad}\eta_{bf}
-2i\eta_{de}\eta_{ac}\eta_{bf}
\right)\right]\nonumber\\
&&\quad\quad=e^{\mu c} e^{\nu d} R_{\mu\nu}^{ab}
\left[\eta_{ac} \eta_{bd}
-4\varphi^{ef}\eta_{ea}\eta_{fc}\eta_{bd}
+4\varphi^{ef}\eta_{ce}\eta_{ad}\eta_{bf}
\right]\nonumber\\
&&\quad\quad=e^{\mu c} e^{\nu d} R_{\mu\nu}^{ab}
\left[\eta_{ac} \eta_{bd}
-4\varphi_{ac}\eta_{bd}
-4\varphi_{bc}\eta_{ad}
\right]\nonumber\\
&&\quad\quad=e^{\mu c} e^{\nu d} R_{\mu\nu}^{ab}
\left[\eta_{ac} \eta_{bd}
+4\varphi_{bc}\eta_{ad}
-4\varphi_{bc}\eta_{ad}
\right]\nonumber\\
&&\quad\quad=e^{\mu c} e^{\nu d} R_{\mu\nu}^{ab}\eta_{ac} \eta_{bd}
=e^\mu_a e^\nu_b R_{\mu\nu}^{ab},
\label{gauge_invariance_Einstein-Hilbert}
\end{eqnarray}
where have been used ($\ref{algebra}$),($\ref{trace_generalized_generators_Lorentz}$)
and ($\ref{transformation_field_strength_sectors}$). Only in the line where the series
expansion is introduced the Landau symbol appears explicitly. The gauge invariance
of the Yang-Mills action on curved space-time is shown in an analogous way,

\begin{eqnarray}
&&\frac{1}{4}e^{\mu a} e^{\nu b} e^\rho_a e^\sigma_b G_{\mu\nu}^A G_{\rho\sigma}^A
=\frac{1}{2}{\rm tr}\left[e^{\mu a} e^{\nu b} e^\rho_a e^\sigma_b G_{\mu\nu}^A T^A 
G_{\rho\sigma}^B T^B\right]\nonumber\\
&&\longrightarrow \frac{1}{2}{\rm tr}\left[e^{\mu a} e^{\nu b} e^\rho_a e^\sigma_b
\mathcal{U}_G \left(G_{\mu\nu}^A T^A\right)\gamma^0 \mathcal{U}_G^{\dagger} \gamma^0\mathcal{U}_G
\left(G_{\rho\sigma}^B T^B\right)\gamma^0 \mathcal{U}_G^{\dagger} \gamma^0\right]
=\frac{1}{2} e^{\mu a} e^{\nu b}e^\rho_a e^\sigma_b
G_{\mu\nu}^A G_{\rho\sigma}^B\ {\rm tr}\left[\mathcal{U}_G \left(T^A T^B\right)
\gamma^0 \mathcal{U}_G^{\dagger} \gamma^0\right]\nonumber\\
&&\quad\quad=\frac{1}{2}e^{\mu a} e^{\nu b}e^\rho_a e^\sigma_b
G_{\mu\nu}^A G_{\rho\sigma}^B\ {\rm tr}\left[\left({\bf 1}+i\alpha^C T^C+i\varphi^{cd}\Sigma_{cd}\right)
\left(T^A T^B\right)\left({\bf 1}-i\alpha^C T^C-i\varphi^{cd}\Sigma_{cd}\right)\right]
+\mathcal{O}\left(\alpha^2,\varphi^2\right)\nonumber\\
&&\quad\quad=\frac{1}{2}e^{\mu a} e^{\nu b}e^\rho_a e^\sigma_b
G_{\mu\nu}^A G_{\rho\sigma}^B\ {\rm tr}\left[T^A T^B+i\alpha^C\left(T^C T^A T^B-T^A T^B T^C\right)
+i\varphi^{cd}\left(\Sigma_{cd}T^A T^B-T^A T^B \Sigma_{cd}\right)\right]\nonumber\\
&&\quad\quad=\frac{1}{2}e^{\mu a} e^{\nu b}e^\rho_a e^\sigma_b
G_{\mu\nu}^A G_{\rho\sigma}^B\ {\rm tr}\left[T^A T^B+i\alpha^C\left(if^{CAD}T^D T^B-if^{BCD}T^A T^D\right)
+i\varphi^{cd}\left(\Sigma_{cd}T^A T^B-T^A T^B \Sigma_{cd}\right)\right]\nonumber\\
&&\quad\quad=\frac{1}{4}e^{\mu a} e^{\nu b} e^\rho_a e^\sigma_b
G_{\mu\nu}^A G_{\rho\sigma}^B\ \left[\delta^{AB}+i\alpha^C\left(if^{CAD}\delta^{DB}-if^{BCD}\delta^{AD}\right)
+i\varphi^{cd}\left(\Sigma_{cd}\delta^{AB}-\delta^{AB}\Sigma_{cd}\right)\right]\nonumber\\
&&\quad\quad=\frac{1}{4}e^{\mu a} e^{\nu b} e^\rho_a e^\sigma_b
G_{\mu\nu}^A G_{\rho\sigma}^B\ \left[\delta^{AB}+i\alpha^C\left(if^{CAB}-if^{BCA}\right)\right]\nonumber\\
&&\quad\quad=\frac{1}{4}e^{\mu a} e^{\nu b} e^\rho_a e^\sigma_b G_{\mu\nu}^A G_{\rho\sigma}^B \delta^{AB}
=\frac{1}{4}e^{\mu a} e^{\nu b} e^\rho_a e^\sigma_b G_{\mu\nu}^A G_{\rho\sigma}^A,
\label{gauge_invariance_Yang-Mills}
\end{eqnarray}
where have been used ($\ref{algebra}$),($\ref{trace_generalized_generators_Yang-Mills}$) and
($\ref{transformation_field_strength_sectors}$). The decisive term is of course the intersection term, which
is considered now with respect to gauge invariance. Within the corresponding calculation it is useful to
express the generalized generators by the usual generators by using ($\ref{definition_generators}$).
The gauge invariance can be shown as follows:

\begin{eqnarray}
&&\frac{1}{4}e^{\mu e}e^{\rho}_{e}e^{\nu f}e^{\sigma}_{f}
\mathcal{H}^{Aab}_{\mu\nu}\mathcal{H}^{A}_{\rho\sigma ab}
=\frac{1}{8}e^{\mu e}e^{\rho}_{e}e^{\nu f}e^{\sigma}_{f}
\mathcal{H}^{Aab}_{\mu\nu}\mathcal{H}^{Acd}_{\rho\sigma}
\left(-\eta_{ab}\eta_{cd}+\eta_{ac}\eta_{bd}
-\eta_{ad}\eta_{bc}\right)
=\frac{1}{4}{\rm tr}\left[e^{\mu e}e^{\nu f}e^{\rho}_{e}e^{\sigma}_{f}
\mathcal{H}^{Aab}_{\mu\nu}\tau^{A}\sigma_{ab}
\mathcal{H}^{Bcd}_{\rho\sigma}
\tau^{B}\sigma_{cd}\right]
\nonumber\\&&\longrightarrow
\frac{1}{4}{\rm tr}\left[e^{\mu e}e^{\nu f}e^{\rho}_{e}e^{\sigma}_{f}
\mathcal{U}_G\left(\mathcal{H}^{Aab}_{\mu\nu}
\tau^{A}\sigma_{ab}\right)\mathcal{U}_G^{\dagger}\mathcal{U}_G\left(\mathcal{H}^{Bcd}_{\rho\sigma}\tau^{B}\sigma_{cd}\right)\mathcal{U}_G^{\dagger}\right]
=\frac{1}{4}e^{\mu e}e^{\nu f}e^{\rho}_{e}e^{\sigma}_{f}\mathcal{H}^{Aab}_{\mu\nu}
\mathcal{H}^{Bcd}_{\rho\sigma}\ {\rm tr}\left[\mathcal{U}_G\left(\tau^{A}\sigma_{ab}
\tau^{B}\sigma_{cd}\right)\mathcal{U}_G^{\dagger}\right]
\nonumber\\&&\quad\quad
=\frac{1}{4}e^{\mu e}e^{\nu f}e^{\rho}_{e}e^{\sigma}_{f}
\mathcal{H}^{Aab}_{\mu\nu}\mathcal{H}^{Bcd}_{\rho\sigma}{\rm tr}
\left[\left({\bf 1}+i\alpha^C T^C+i\varphi^{ef}\Sigma_{ef}\right)
\left(\tau^{A}\sigma_{ab}\tau^{B}\sigma_{cd}\right)
\left({\bf 1}-i\alpha^C T^C-i\varphi^{ef}\Sigma_{ef}\right)\right]
+\mathcal{O}\left(\alpha^2,\varphi^2\right)
\nonumber\\&&\quad\quad
=\frac{1}{4}e^{\mu e}e^{\nu f}e^{\rho}_{e}e^{\sigma}_{f}\mathcal{H}^{Aab}_{\mu\nu}
\mathcal{H}^{Bcd}_{\rho\sigma}\ {\rm tr}\left[\left({\bf 1}+i\alpha^C \chi^{CD}\tau^D
+i\varphi^{ef}M_{ef}^{\ \ gh}
\sigma_{gh}\right)\left(\tau^{A}\sigma_{ab}\tau^{B}\sigma_{cd}\right)
\left({\bf 1}-i\alpha^C\chi^{CD}\tau^D-i\varphi^{ef}M_{ef}^{\ \ gh}
\sigma_{gh}\right)\right]
\nonumber\\&&\quad\quad
=\frac{1}{4}e^{\mu e}e^{\nu f}e^{\rho}_{e}e^{\sigma}_{f}
\mathcal{H}^{Aab}_{\mu\nu}\mathcal{H}^{Bcd}_{\rho\sigma}
\ {\rm tr}\left[\tau^{A}\sigma_{ab}\tau^{B}\sigma_{cd}
+i\alpha^C \chi^{CD}\left(\tau^{D}\tau^{A}\sigma_{ab}\tau^{B}\sigma_{cd}
-\tau^{A}\sigma_{ab}\tau^{B}\sigma_{cd}\tau^{D}\right)
\right.\nonumber\\ &&\left. \quad\quad\quad\quad\quad\quad\quad\quad\quad\quad\quad\quad\quad\quad
+i\varphi^{ef}M_{ef}^{\ \ gh}\left(\sigma_{gh}\tau^{A}\sigma_{ab}
\tau^{B}\sigma_{cd}-\tau^{A}\sigma_{ab}\tau^{B}\sigma_{cd}\sigma_{gh}\right)\right]
\nonumber\\&&\quad\quad
=\frac{1}{4}e^{\mu e}e^{\nu f}e^{\rho}_{e}e^{\sigma}_{f}
\mathcal{H}^{Aab}_{\mu\nu}\mathcal{H}^{Bcd}_{\rho\sigma}
\ {\rm tr}\left[\tau^{A}\sigma_{ab}\tau^{B}\sigma_{cd}
+i\alpha^C \chi^{CD}\sigma_{ab}\sigma_{cd}
\left(\tau^{d}\tau^{A}\tau^{B}-\tau^{A}\tau^{B}\tau^{D}\right)
\right.\nonumber\\&&\quad\quad\quad\quad\quad\quad\quad\quad\quad\quad\quad\quad\quad\quad\left.
+i\varphi^{ef}M_{ef}^{\ \ gh}
\tau^{A}\tau^{B}\left(\sigma_{gh}\sigma_{ab}\sigma_{cd}
-\sigma_{ab}\sigma_{cd}\sigma_{gh}\right)\right]
\nonumber\\&&\quad\quad
=\frac{1}{4}e^{\mu e}e^{\nu f}e^{\rho}_{e}e^{\sigma}_{f}
\mathcal{H}^{Aab}_{\mu\nu}\mathcal{H}^{Bcd}_{\rho\sigma}
\ {\rm tr}\left[\tau^{A}\sigma_{ab}\tau^{B}\sigma_{cd}
+i\alpha^C \chi^{CD}\sigma_{ab}\sigma_{cd}
\left(if^{DAE}\tau^{E}\tau^{B}-if^{BDE}\tau^{A}\tau^{E}\right)
\right.\nonumber\\&&\quad\quad\quad\quad\quad\quad\quad\quad\quad\quad\quad\quad\quad\quad\left.
+i\varphi^{ef}M_{ef}^{\ \ gh}\tau^{A}\tau^{B}
\left(i\eta_{ga}\sigma_{hb}-i\eta_{ha}\sigma_{gb}
-i\eta_{gb}\sigma_{ha}+i\eta_{hb}\sigma_{ga}\right)\sigma_{cd}
\right.\nonumber\\&&\quad\quad\quad\quad\quad\quad\quad\quad\quad\quad\quad\quad\quad\quad\left.
-i\varphi^{ef}M_{ef}^{\ \ gh}\tau^{A}\tau^{B}\sigma_{ab}
\left(i\eta_{cg}\sigma_{dh}-i\eta_{dg}\sigma_{ch}
-i\eta_{ch}\sigma_{dg}+i\eta_{dh}\sigma_{cg}\right)\right]
\nonumber\\&&\quad\quad
=\frac{1}{4}e^{\mu e}e^{\nu f}e^{\rho}_{e}e^{\sigma}_{f}\mathcal{H}^{Aab}_{\mu\nu}
\mathcal{H}^{Bcd}_{\rho\sigma}
\ {\rm tr}\left[\tau^{A}\sigma_{ab}\tau^{B}\sigma_{cd}
+i\alpha^C \chi^{CD}\sigma_{ab}\sigma_{cd}
\left(if^{DAE}\tau^{E}\tau^{B}-if^{BDE}\tau^{A}\tau^{E}\right)
\right.\nonumber\\&&\quad\quad\quad\quad\quad\quad\quad\quad\quad\quad\quad\quad\quad\quad\left.
+i\varphi^{ef}M_{ef}^{\ \ gh}\tau^{A}\tau^{B}
\left(2i\eta_{ga}\sigma_{hb}-2i\eta_{gb}\sigma_{ha}\right)\sigma_{cd}
-i\varphi^{ef}M_{ef}^{\ \ gh}\tau^{A}\tau^{B}\sigma_{ab}
\left(2i\eta_{cg}\sigma_{dh}-2i\eta_{dg}\sigma_{ch}\right)\right]
\nonumber\\&&\quad\quad
=\frac{1}{8}e^{\mu e}e^{\nu f}e^{\rho}_{e}e^{\sigma}_{f}
\mathcal{H}^{Aab}_{\mu\nu}\mathcal{H}^{Bcd}_{\rho\sigma}
\left[\delta^{AB}\left(-\eta_{ab}\eta_{cd}
+\eta_{ac}\eta_{bd}
-\eta_{ad}\eta_{bc}\right)
\right.\nonumber\\&&\quad\quad\quad\quad\quad\quad\quad\quad\quad\quad\quad\quad\quad\quad\left.
+i\alpha^C \chi^{CD}
\left(-\eta_{ab}\eta_{cd}
+\eta_{ac}\eta_{bd}
-\eta_{ad}\eta_{bc}\right)
\left(if^{DAE}\delta^{EB}-if^{BDE}\delta^{AE}\right)
\right.\nonumber\\&&\quad\quad\quad\quad\quad\quad\quad\quad\quad\quad\quad\quad\quad\quad\left.
+i\varphi^{ef}M_{ef}^{\ \ gh}\delta^{AB}
\left(2i\eta_{ga}\left(-\eta_{hb}\eta_{cd}
+\eta_{hc}\eta_{bd}
-\eta_{hd}\eta_{bc}\right)
-2i\eta_{fb}\left(-\eta_{ha}\eta_{cd}
+\eta_{hc}\eta_{ad}
-\eta_{hd}\eta_{ac}\right)\right)
\right.\nonumber\\&&\quad\quad\quad\quad\quad\quad\quad\quad\quad\quad\quad\quad\quad\quad\left.
-i\varphi^{ef}M_{ef}^{\ \ gh}\delta^{AB}
\left(2i\eta_{cg}\left(-\eta_{ab}\eta_{dh}
+\eta_{ad}\eta_{bh}
-\eta_{ah}\eta_{bd}\right)
-2i\eta_{di}\left(-\eta_{ab}\eta_{ch}
+\eta_{ac}\eta_{bh}
-\eta_{ah}\eta_{bc}\right)
\right)\right]
\nonumber\\&&\quad\quad
=\frac{1}{8}e^{\mu e}e^{\nu f}e^{\rho}_{e}e^{\sigma}_{f}
\mathcal{H}^{Aab}_{\mu\nu}\mathcal{H}^{Bcd}_{\rho\sigma}
\left[2\delta^{AB}\eta_{ac}\eta_{bd}
+2i\alpha^C \chi^{CD}\eta_{ac}\eta_{bd}
\left(if^{DAE}\delta^{EB}-if^{BDE}\delta^{AE}\right)
\right.\nonumber\\&&\quad\quad\quad\quad\quad\quad\quad\quad\quad\quad\quad\quad\quad\quad\left.
+i\varphi^{ef}M_{ef}^{\ \ gh}\delta^{AB}
\left(4i\eta_{ga}\eta_{hc}\eta_{bd}
-4i\eta_{gb}\eta_{hc}\eta_{ad}\right)
\right.\nonumber\\&&\quad\quad\quad\quad\quad\quad\quad\quad\quad\quad\quad\quad\quad\quad\left.
-i\varphi^{ef}M_{ef}^{\ \ gh}\delta^{AB}
\left(4i\eta_{cg}\eta_{ad}\eta_{bh}
-4i\eta_{dg}\eta_{ac}\eta_{bh}\right)\right]
\nonumber\\&&\quad\quad
=\frac{1}{8}e^{\mu e}e^{\nu f}e^{\rho}_{e}e^{\sigma}_{f}
\mathcal{H}^{Aab}_{\mu\nu}\mathcal{H}^{Bcd}_{\rho\sigma}
\left[2\delta^{AB}\eta_{ac}\eta_{bd}
+2i\alpha^C \chi^{CD}\eta_{ac}\eta_{bd}
\left(if^{DAB}-if^{BDA}\right)
\right.\nonumber\\&&\quad\quad\quad\quad\quad\quad\quad\quad\quad\quad\quad\quad\quad\quad\left.
-8\varphi^{ef}M_{ef}^{\ \ gh}\delta^{AB}
\eta_{ga}\eta_{hc}\eta_{bd}
+8\varphi^{ef}M_{ef}^{\ \ gh}\delta^{AB}
\eta_{cg}\eta_{ad}\eta_{bh}\right]
\nonumber\\&&\quad\quad
=\frac{1}{8}e^{\mu e}e^{\nu f}e^{\rho}_{e}e^{\sigma}_{f}
\mathcal{H}^{Aab}_{\mu\nu}\mathcal{H}^{Bcd}_{\rho\sigma}
\left[2\delta^{AB}\eta_{ac}\eta_{bd}
-8\varphi^{ef}M_{ef ac}\delta^{AB}\eta_{bd}
+8\varphi^{ef}M_{ef cb}\delta^{AB}\eta_{ad}\right]
\nonumber\\&&\quad\quad
=\frac{1}{8}e^{\mu e}e^{\nu f}e^{\rho}_{e}e^{\sigma}_{f}
\mathcal{H}^{Aab}_{\mu\nu}\mathcal{H}^{Bcd}_{\rho\sigma}
\left[2\delta^{AB}\eta_{ac}\eta_{bd}
-8\varphi^{ef}M_{ef ac}\delta^{AB}\eta_{bd}
-8\varphi^{ef}M_{ef bc}\delta^{AB}\eta_{ad}\right]
\nonumber\\&&\quad\quad
=\frac{1}{8}e^{\mu e}e^{\nu f}e^{\rho}_{e}e^{\sigma}_{f}
\mathcal{H}^{Aab}_{\mu\nu}\mathcal{H}^{Bcd}_{\rho\sigma}
\left[2\delta^{AB}\eta_{ac}\eta_{bc}
+8\varphi^{ef}M_{efac}\delta^{AB}\eta_{bd}
-8\varphi^{ef}M_{efac}\delta^{AB}\eta_{bd}\right]
\nonumber\\&&\quad\quad
=\frac{1}{4}e^{\mu e}e^{\nu f}e^{\rho}_{e}e^{\sigma}_{f}\mathcal{H}^{Aab}_{\mu\nu}
\mathcal{H}^{Bcd}_{\rho\sigma}\delta^{AB}\eta_{ac}\eta_{bc}
\nonumber\\&&\quad\quad
=\frac{1}{4}e^{\mu e}e^{\nu f}e^{\rho}_{e}e^{\sigma}_{f}\mathcal{H}^{Aab}_{\mu\nu}
\mathcal{H}^{A}_{\rho\sigma ab},
\label{gauge_invariance_intersection}
\end{eqnarray}
where have been used ($\ref{antisymmetry}$),($\ref{trace_usual_generators}$) and ($\ref{transformation_field_strength_sectors}$)
as well as the property $\mathcal{H}^{Aab}_{\mu\nu}\eta_{ab}=0$, which is valid since $\mathcal{H}^{Aab}_{\mu\nu}$ is
antisymmetric with respect to the Lorentz indices,
$\mathcal{H}^{Aab}_{\mu\nu}=-\mathcal{H}^{Aba}_{\mu\nu}$. In ($\ref{gauge_invariance_intersection}$) the trace refers
independently to the generators of the $SU(N)$ group as well as to the generators of the $SO(3,1)$ group, whereas
in ($\ref{gauge_invariance_Yang-Mills}$) it refers only to the generators of the $SU(N)$ group and in
($\ref{gauge_invariance_Einstein-Hilbert}$) it refers only to the generators of the $SO(3,1)$ group.
Thus it has been shown that the generalized combined action for Yang-Mills theory and general relativity considered
according to ($\ref{action_gauge_fields}$) is indeed gauge invariant under the generalized gauge group based on the
algebra ($\ref{algebra}$) and mediated by the transformation operator ($\ref{transformation_operator}$) and can thus
be postulated as action of the intersection gauge theory.

\section{Generalized Energy Momentum Tensor and Corresponding Einstein Field Equation}

To obtain the generalized Einstein field equation, the complete action incorporating the fermionic matter action $\mathcal{S}_M$ corresponding to the Lagrangian ($\ref{gauge_invariant_matter_Lagrangian}$) as well as the generalized
action of the gauge fields $\mathcal{S}_G$, $\mathcal{S}=\mathcal{S}_M+\mathcal{S}_G$, has to be varied with respect
to the tetrad field $e^\mu_a$.
Since the action of the gravitational field itself without any interaction with the Yang-Mills field is not changed, the left
hand side of the Einstein field equation remains also unmodified. Because of the additional term containing the intersection
field strength $\mathcal{H}^{Aab}_{\mu\nu}$ the energy momentum tensor with respect to the Yang-Mills field is however
changed decisively. This means that the general definition of the energy momentum tensor with respect to the complete action
containing matter fields has to to be considered, 

\begin{equation}
\mathcal{T}_{\mu}^{a}=-\frac{1}{e}\frac{\delta\left(\mathcal{S}_M+\mathcal{S}_{YM}
+\mathcal{S}_{Int}\right)}{\delta e^\mu_a},
\label{energy-momentum_tensor_definition}
\end{equation}
Since the action of the fermionic matter field $\mathcal{S}_M$ as well as the action of the Yang-Mills field
$\mathcal{S}_{YM}$ are equivalent to the corresponding usual actions formulated on curved space-time
respectively, their energy momentum tensors are not written explicitly here and accordingly it is made
the following definition:

\begin{equation}
\mathcal{T}_{M\ \mu}^{\ \ \ a}=-\frac{1}{e}\frac{\delta \mathcal{S}_{M}}{\delta e^\mu_a},\quad
\mathcal{T}_{YM\ \mu}^{\ \ \ \ \ a}=-\frac{1}{e}\frac{\delta \mathcal{S}_{YM}}{\delta e^\mu_a}.
\label{usual_energy-momentum_tensors}
\end{equation}
The decisive term within the energy momentum tensor ($\ref{energy-momentum_tensor_definition}$) is of course the
term induced by the intersection action $\mathcal{S}_{Int}$, which is built from the new sector of the field
strength. It is obtained by varying this action,

\begin{equation}
\delta \mathcal{S}_{Int}=\frac{1}{4}\int d^4 x\ \delta\left(e e^{\mu a} e^\rho_a e^{\nu b} e^\sigma_b
\mathcal{H}^{Acd}_{\mu\nu} \mathcal{H}^{A}_{\rho\sigma cd}\right).
\label{variation_intersection_action}
\end{equation}
Variation of the expression for the Lagrangian within the integral yields the following expression:

\begin{eqnarray}
\delta\left(e e^{\mu a} e^\rho_a e^{\nu b} e^\sigma_b \mathcal{H}_{\mu\nu}^{Acd} \mathcal{H}_{\rho\sigma cd}^{A}\right)
&=&\delta e e^{\mu a} e^\rho_a e^{\nu b} e^\sigma_b \mathcal{H}_{\mu\nu}^{Acd} \mathcal{H}_{\rho\sigma cd}^{A}
+4e \delta e^{\mu a} e^\rho_a e^{\nu b} e^\sigma_b \mathcal{H}_{\mu\nu}^{Acd} \mathcal{H}_{\rho\sigma cd}^{A}
+2 e e^{\mu a} e^\rho_a e^{\nu b} e^\sigma_b \delta \mathcal{H}_{\mu\nu}^{Acd} \mathcal{H}_{\rho\sigma cd}^{A}
\nonumber\\
&=&-e e^c_\lambda \delta e^\lambda_c e^{\mu a} e^\rho_a e^{\nu b}
e^\sigma_b \mathcal{H}_{\mu\nu}^{Acd} \mathcal{H}_{\rho\sigma cd}^{A}
+4e \delta e^{\mu a} e^\rho_a e^{\nu b} e^\sigma_b \mathcal{H}_{\mu\nu}^{A cd} \mathcal{H}_{\rho\sigma cd}^{A}
\nonumber\\
&&+2 e e^{\mu a} e^\rho_a e^{\nu b} e^\sigma_b \left(A_\mu^B \delta \omega_\nu^{ef}-A_\nu^B \delta \omega_\mu^{ef}\right) \Gamma_{ef}^{BAcd} \mathcal{H}_{\rho\sigma cd}^{A}
\nonumber\\
&=&-e e^c_\lambda \delta e^\lambda_c e^{\mu a} e^\rho_a e^{\nu b}
e^\sigma_b \mathcal{H}_{\mu\nu}^A \mathcal{H}_{\rho\sigma cd}^{A}
+4e \delta e^{\mu a} e^\rho_a e^{\nu b} e^\sigma_b \mathcal{H}_{\mu\nu}^{Acd} \mathcal{H}_{\rho\sigma cd}^{A}
+4 e e^{\mu a} e^\rho_a e^{\nu b} e^\sigma_b A_\mu^B \Gamma_{ef}^{BAcd} \mathcal{H}_{\rho\sigma cd}^{A}\delta \omega_\nu^{ef}
\nonumber\\
&=&-e e^c_\lambda \delta e^\lambda_c e^{\mu a} e^\rho_a e^{\nu b}
e^\sigma_b \mathcal{H}_{\mu\nu}^{Acd} \mathcal{H}_{\rho\sigma cd}^{A}
+4e \delta e^{\mu a} e^\rho_a e^{\nu b} e^\sigma_b \mathcal{H}_{\mu\nu}^{Acd} \mathcal{H}_{\rho\sigma cd}^{A}
+4 \Gamma_{ef}^{BAcd}\Xi^{ABefg}_{cd\lambda}\left(A,e\right)
\delta e^\lambda_g\nonumber\\
&=&\left[-e e^g_\lambda e^{\mu a} e^\rho_a e^{\nu b}e^\sigma_b
\mathcal{H}_{\mu\nu}^{Acd} \mathcal{H}_{\rho\sigma cd}^{A}
+4e \delta^\mu_\lambda \delta^{ag} e^\rho_a e^{\nu b} e^\sigma_b \mathcal{H}_{\mu\nu}^{Acd} \mathcal{H}_{\rho\sigma cd}^{A}
+4\Gamma_{ef}^{BAcd}
\Xi^{ABefg}_{cd\lambda}\left(A,e\right)\right]
\delta e^\lambda_g,
\label{variation_intersection_Lagrangian}
\end{eqnarray}
where has been used the variation rule of the determinant of the tetrad field,
$\delta e=-e e_\mu^a \delta e^\mu_a$,
($\ref{reformulation_intersection_field_strength}$) and accordingly

\begin{eqnarray}
\delta \mathcal{H}_{\mu\nu}^{Aab}&=&\delta H_{\mu\nu}^{Bcd}\Gamma_{\rho\sigma}^{BAab}
=\left(A_\mu^B \delta \omega_\nu^{cd}-A_\nu^B \delta \omega_\mu^{cd}\right)\Gamma_{cd}^{BAab},
\end{eqnarray}
and a relation, which is derived in the following calculation:

\begin{eqnarray}
\int d^4 x\ f \delta \omega_\mu^{ab}&=&\int d^4 x\ f
\left(2\delta e^{\nu a}\partial_\mu e_\nu^b
+2e^{\nu a}\partial_\mu \delta e_\nu^b
-2\delta e^{\nu b} \partial_\mu e_\nu^a
-2e^{\nu b} \partial_\mu \delta e_\nu^a
\right.\nonumber\\&&\left.
-2\delta e^{\nu a}\partial_\nu e_\mu^b
-2e^{\nu a}\partial_\nu \delta e_\mu^b
+2\delta e^{\nu b} \partial_\nu e_\mu^a
+2e^{\nu b} \partial_\nu \delta e_\mu^a
\right.\nonumber\\&&\left.
+\delta e_{\mu c} e^{\nu a}e^{\sigma b}\partial_\sigma e_\nu^c
+e_{\mu c} \delta e^{\nu a}e^{\sigma b}\partial_\sigma e_\nu^c
+e_{\mu c} e^{\nu a} \delta e^{\sigma b}\partial_\sigma e_\nu^c
+e_{\mu c} e^{\nu a}e^{\sigma b}\partial_\sigma \delta e_\nu^c
\right.\nonumber\\&&\left.
-\delta e_{\mu c} e^{\nu a}e^{\sigma b}\partial_\nu e_\sigma^c
-e_{\mu c} \delta e^{\nu a}e^{\sigma b}\partial_\nu e_\sigma^c
-e_{\mu c} e^{\nu a} \delta e^{\sigma b}\partial_\nu e_\sigma^c
-e_{\mu c} e^{\nu a}e^{\sigma b}\partial_\nu \delta e_\sigma^c\right)\nonumber\\
&=&\int d^4 x\ \left[2f \delta^{\nu}_\rho \delta^{ad}\partial_\mu e_\nu^b
-2\delta_{\nu\rho}\delta^{bd}\partial_\mu\left(f e^{\nu a}\right)
-2f\delta^{\nu}_\rho \delta^{bd} \partial_\mu e_\nu^a
+2\delta_{\nu\rho}\delta^{ad}\partial_\mu\left(f e^{\nu b}\right)
\right.\nonumber\\&&\left.
-2f\delta^{\nu}_\rho \delta^{ad}\partial_\nu e_\mu^b
+2\delta_{\mu\rho}\delta^{bd}\partial_\nu\left(f e^{\nu a}\right)
+2f\delta^\nu_\rho \delta^{bd}\partial_\nu e_\mu^a
-2\delta_{\mu\rho}\delta^{ad}\partial_\nu \left(f e^{\nu b}\right)
\right.\nonumber\\&&\left.
+f\delta_{\mu\rho} \delta^d_c e^{\nu a}e^{\sigma b}\partial_\sigma e_\nu^c
+f e_{\mu c} \delta^{\nu}_\rho \delta^{ad} e^{\sigma b}\partial_\sigma e_\nu^c
+f e_{\mu c} e^{\nu a} \delta^\sigma_\rho \delta^{bd}\partial_\sigma e_\nu^c
-\delta_{\nu\rho}\delta^{cd}\partial_\sigma\left(f e_{\mu c} e^{\nu a}e^{\sigma b}\right)
\right.\nonumber\\&&\left.
-f\delta_{\mu\rho} \delta^{d}_{c} e^{\nu a}e^{\sigma b}\partial_\nu e_\sigma^c
-f e_{\mu c} \delta^{\nu}_{\rho} \delta^{ad} e^{\sigma b}\partial_\nu e_\sigma^c
-f e_{\mu c} e^{\nu a} \delta^{\sigma}_\rho \delta^{bd}\partial_\nu e_\sigma^c
+\delta_{\sigma\rho}\delta^{cd}\partial_\nu \left(f e_{\mu c} e^{\nu a}e^{\sigma b}\right)\right]
\delta e^\rho_d,%\nonumber\\
\end{eqnarray}
where $f$ denotes an arbitrary function and which becomes manifest with respect to
($\ref{variation_intersection_Lagrangian}$) in the following special form: 

\begin{eqnarray}
\int d^4 x\ e e^{\mu a} e^\rho_a e^{\nu b} e^\sigma_b \mathcal{H}_{\rho\sigma cd}^{A}
A_\mu^B \delta \omega_\nu^{ef}
&=&\int d^4 x\ \left[2e e^{\mu a} e^\rho_a e^{\nu b} e^\sigma_b \mathcal{H}_{\rho\sigma cd}^{A} A_\mu^B \delta^{g}_\lambda \delta^{ei}\partial_\nu e_g^f
-2\delta_{g\lambda}\delta^{fi}\partial_\nu\left(e e^{\mu a} e^\rho_a e^{\nu b} e^\sigma_b \mathcal{H}_{\rho\sigma cd}^{A}
A_\mu^B e^{ge}\right)
\right.\nonumber\\&&\left.
-2e e^{\mu a} e^\rho_a e^{\nu b} e^\sigma_b \mathcal{H}_{\rho\sigma cd}^{A} A_\mu^B\delta^{g}_\lambda \delta^{fi} \partial_\nu e_g^e
+2\delta_{g\lambda}\delta^{ei}\partial_\nu\left(e e^{\mu a} e^\rho_a e^{\nu b} e^\sigma_b \mathcal{H}_{\rho\sigma cd}^{A} A_\mu^B e^{gf}\right)
\right.\nonumber\\&&\left.
-2e e^{\mu a} e^\rho_a e^{\nu b} e^\sigma_b \mathcal{H}_{\rho\sigma cd}^{A} A_\mu^B\delta^{g}_\lambda \delta^{ei}\partial_g e_\nu^f
+2\delta_{\nu\lambda}\delta^{fi}\partial_g\left(e e^{\mu a} e^\rho_a e^{\nu b} e^\sigma_b \mathcal{H}_{\rho\sigma cd}^{A} A_\mu^B e^{ge}\right)
\right.\nonumber\\&&\left.
+2e e^{\mu a} e^\rho_a e^{\nu b} e^\sigma_b \mathcal{H}_{\rho\sigma cd}^{A} A_\mu^B\delta^g_\lambda \delta^{fi}\partial_g e_\nu^e
-2\delta_{\nu\lambda}\delta^{ei}\partial_g \left(e e^{\mu a} e^\rho_a e^{\nu b} e^\sigma_b \mathcal{H}_{\rho\sigma cd}^{A} A_\mu^B e^{gf}\right)
\right.\nonumber\\&&\left.
+e e^{\mu a} e^\rho_a e^{\nu b} e^\sigma_b \mathcal{H}_{\rho\sigma cd}^{A} A_\mu^B\delta_{\nu\lambda} \delta^i_h e^{ge}e^{\sigma f}\partial_\sigma e_g^h
+e e^{\mu a} e^\rho_a e^{\nu b} e^\sigma_b \mathcal{H}_{\rho\sigma cd}^{A} A_\mu^B e_{\nu h} \delta^{g}_\lambda \delta^{ei} e^{\sigma f}\partial_\sigma e_g^h
\right.\nonumber\\&&\left.
+e e^{\mu a} e^\rho_a e^{\nu b} e^\sigma_b \mathcal{H}_{\rho\sigma cd}^{A} A_\mu^B e_{\nu h} e^{ge} \delta^\sigma_\lambda \delta^{fi}\partial_\sigma e_g^h
-\delta_{g\lambda}\delta^{hi}\partial_\sigma\left(e e^{\mu a} e^\rho_a e^{\nu b} e^\sigma_b \mathcal{H}_{\rho\sigma cd}^{A} A_\mu^B e_{\nu h} e^{ge}e^{\sigma f}\right)
\right.\nonumber\\&&\left.
-e e^{\mu a} e^\rho_a e^{\nu b} e^\sigma_b \mathcal{H}_{\rho\sigma cd}^{A} A_\mu^B\delta_{\nu\lambda} \delta^i_h e^{ge}e^{\sigma f}\partial_g e_\sigma^h
-e e^{\mu a} e^\rho_a e^{\nu b} e^\sigma_b \mathcal{H}_{\rho\sigma cd}^{A} A_\mu^B e_{\nu h} \delta^{g}_{\lambda} \delta^{ei} e^{\sigma f}\partial_g e_\sigma^h
\right.\nonumber\\&&\left.
-e e^{\mu a} e^\rho_a e^{\nu b} e^\sigma_b \mathcal{H}_{\rho\sigma cd}^{A} A_\mu^B e_{\nu h} e^{ge} \delta^{\sigma}_\lambda \delta^{fi}\partial_g e_\sigma^h
+\delta_{\sigma\lambda}\delta^{hi}\partial_g \left(e e^{\mu a} e^\rho_a e^{\nu b} e^\sigma_b \mathcal{H}_{\rho\sigma cd}^{A} A_\mu^B e_{\nu h} e^{ge}e^{\sigma f}\right)\right]
\delta e^\lambda_i\nonumber\\
&\equiv&\int d^4 x\ \Xi^{ABefi}_{cd\lambda}\left(A,e\right) \delta e^{\lambda}_i.
\label{variation_omega_definition_Xi}
\end{eqnarray}
The last line of ($\ref{variation_omega_definition_Xi}$) represents a definition of the new quantity $\Xi^{Abefi}_{cd\lambda}\left(A,e\right)$, which depends accordingly on the Yang-Mills
field and the gravitational field.
Inserting ($\ref{usual_energy-momentum_tensors}$) and ($\ref{variation_intersection_action}$) with
($\ref{variation_intersection_Lagrangian}$) into the definition of the generalized energy momentum tensor
($\ref{energy-momentum_tensor_definition}$) yields the following expression for the generalized energy
momentum tensor corresponding to the generalization of the action of the Yang-Mills field and the
gravitational field according to ($\ref{action_gauge_fields}$) as invariant action under
gauge transformations induced by the generalized algebra ($\ref{algebra}$):

\begin{eqnarray}
\mathcal{T}_{\lambda}^{g}&=&\mathcal{T}_{M\ \lambda}^{\ \ \ g}
+\mathcal{T}_{YM\ \lambda}^{\ \ \ \ \ g}-\frac{1}{4e}
\frac{\delta(e e^{\mu a} e^\rho_a e^{\nu b} e^\sigma_b
\mathcal{H}_{\mu\nu}^{Acd} \mathcal{H}_{\rho\sigma cd}^{A})}{\delta e^\lambda_g}
\nonumber\\
&=&\mathcal{T}_{M\ \lambda}^{\ \ \ g}+\mathcal{T}_{YM\ \lambda}^{\ \ \ \ \ c}-\frac{1}{4e}
\left[-e e^g_\lambda e^{\mu a} e^\rho_a e^{\nu b}e^\sigma_b
\mathcal{H}_{\mu\nu}^{Acd} \mathcal{H}_{\rho\sigma cd}^{A}
+4e \delta^\mu_\lambda \delta^{ag} e^\rho_a e^{\nu b} e^\sigma_b \mathcal{H}_{\mu\nu}^{Acd}
\mathcal{H}_{\rho\sigma cd}^{A}+4\Gamma_{ef}^{BAcd}
\Xi^{ABefg}_{cd\lambda}\left(A,e\right)\right]
\nonumber\\
&=&\mathcal{T}_{M\ \lambda}^{\ \ \ g}
+\mathcal{T}_{YM\ \lambda}^{\ \ \ \ \ g}
+\left[\frac{1}{4}e^g_\lambda e^{\mu a} e^\rho_a e^{\nu b}e^\sigma_b
\mathcal{H}_{\mu\nu}^{Acd} \mathcal{H}_{\rho\sigma cd}^{A}
-\delta^\mu_\lambda \delta^{ag} e^\rho_a e^{\nu b} e^\sigma_b \mathcal{H}_{\mu\nu}^{Acd}
\mathcal{H}_{\rho\sigma cd}^{A}-\frac{1}{e}\Gamma_{ef}^{BAcd}
\Xi^{ABefg}_{cd\lambda}\left(A,e\right)\right].
\label{generalized_energy_momentum_tensor}
\end{eqnarray}
The generalized Einstein field equation,
$\frac{1}{e}\frac{\delta \mathcal{S}_{EH}}{\delta e^\mu_a}
=-\frac{1}{e}\frac{\delta\left(\mathcal{S}_{M}+\mathcal{S}_{YM}+\mathcal{S}_{Int}\right)}
{\delta e^\mu_a}=\mathcal{T}_\mu^{a}$, can now be written explicitly. 
As already mentioned, the left hand side of the corresponding generalized Einstein field equation is not modified,
since the Einstein-Hilbert action as action of the free gravitational field according to ($\ref{action_gauge_fields}$)
remains unchanged implying $R_\mu^a-\frac{1}{2}R e_\mu^{a}=8\pi G \mathcal{T}_\mu^{a}$.
Therefore the generalized Einstein field equation is obtained by inserting the generalized energy momentum tensor
containing the intersection term ($\ref{generalized_energy_momentum_tensor}$) to the usual form of the Einstein
field equation leading to
 
\begin{eqnarray}
R_\lambda^g-\frac{1}{2}R e_\lambda^{g}&=&8\pi G\left[\mathcal{T}_{M\ \lambda}^{\ \ \ g}
+\mathcal{T}_{YM\ \lambda}^{\ \ \ \ \ g}
+\frac{1}{4}e^g_\lambda e^{\mu a} e^\rho_a e^{\nu b}e^\sigma_b
\mathcal{H}_{\mu\nu cd}^{A} \mathcal{H}_{\rho\sigma}^{Acd}
-\delta^\mu_\lambda \delta^{ag} e^\rho_a e^{\nu b} e^\sigma_b \mathcal{H}_{\mu\nu}^{Acd}
\mathcal{H}_{\rho\sigma cd}^{A}-\frac{1}{e}\Gamma_{ef}^{BAcd}
\Xi^{ABefg}_{cd\lambda}\left(A,e\right)\right],\nonumber\\
\end{eqnarray}
where $R_\mu^a$ denotes the Ricci tensor, which is defined as $R_\mu^a=e^\nu_b R_{\mu\nu}^{ab}$, and $R$ denotes the Ricci
scalar, which is defined as $R=e^\mu_a e^\nu_b R_{\mu\nu}^{ab}$. The generalized field equation of the Yang-Mills field
is obtained by varying the generalized action ($\ref{action_gauge_fields}$) with respect to the Yang-Mills field
$A_\mu^A$ and thus reads as follows:

\begin{eqnarray}
\frac{\delta\left(\mathcal{S}_{YM}+\mathcal{S}_M+\mathcal{S}_{Int}\right)}{\delta A_\sigma^E}
&=&e e^{\mu a}e^\rho_a e^{\nu b}e^\sigma_b
\left[-\partial_\mu\left(\partial_\nu A_\rho^{E}-\partial_\rho A_\nu^{E}\right)
-\left(\partial_\mu A_\nu^A-\partial_\nu A_\mu^A\right)f^{ABE}A_\rho^B
+f^{ABC}f^{ADE}A_\mu^B A_\nu^C A_\rho^D\right]
\nonumber\\
&&-e\bar \psi \gamma^m e_m^\sigma T^E \psi
+e e^{\mu a}e^\rho_a e^{\nu b}e^\sigma_b \left[A_\mu^B \omega_\nu^{cd}\omega_\rho^{ef}
-A_\nu^B \omega_\mu^{cd}\omega_\rho^{ef}\right]\Gamma^{BAij}_{cd} \Gamma^{EAij}_{ef}=0.
\end{eqnarray}
This interaction structure of the Yang-Mills field with the gravitational field is much more complicated compared with
the usual case and accordingly represents a violation of the equivalence principle.

\section{Summary and Discussion}

It has been considered an intersection of Yang-Mills theory with the gauge description of general relativity.
This intersection has its origin in a generalized algebra, where the generators of the internal $SU(N)$ group
and the generators of the external $SO(3,1)$ group fulfil nontrivial commutation relations.
The commutator is assumed to be equal to an element in the tensor space of the space of the usual generators of the
$SU(N)$ group and the space of the usual generators of the $SO(3,1)$ group, $SU(N) \otimes SO(3,1)$.
This generalized algebra can be realized by building quantities, which are linear combinations of the usual $SU(N)$
generators and the usual $SO(3,1)$ generators respectively under the assumption that the coefficients of the $SU(N)$
generators do not commute with the coefficients of the $SO(3,1)$ generators. 
Under special conditions on the coefficients the generalized quantities fulfil the same algebraic properties
as the usual generators except that they obey the additional commutation relations. If local gauge invariance
of a matter field equation is postulated with respect to the gauge group corresponding to this generalized
algebra, this leads to a generalization of the combination of Yang-Mills gauge theory and the $SO(3,1)$
gauge description of general relativity. Within this generalization all quantities are influenced by the complete
gauge group, since in contrast to the usual case the two gauge groups are not independent of each other anymore.
This implies extended transformation rules for the gauge fields. The dynamics of fermionic matter is not
influenced by the generalization of the algebra, since the usual local gauge invariant Dirac Lagrangian
remains invariant, if there are considered the generalized gauge transformations.
But if the corresponding field strength tensor is built from the covariant derivative containing the generalized
generators, an additional sector appears, which consists of the connection of the Yang-Mills field as well as the
connection of general relativity. From this intersection field strength a new sector of the action has been
constructed in analogy to the Yang-Mills action, which is therefore quadratic in the new field strength.
It has been shown that not only this new action but also the usual Einstein-Hilbert action as well as the usual
Yang-Mills action in curved space-time are still invariant under the generalized gauge transformations. After this,
the energy momentum tensor under incorporation of the new dynamical term has been calculated to obtain the
corresponding generalized inhomogeneous Einstein field equation.
The field equation of the Yang-Mills field are of course also modified decisively and this leads to a violation
of the equivalence principle.
 
Since the presented theory establishes a relation between the internal $SU(N)$ symmetry referring to a quantum
number and the external $SO(3,1)$ symmetry, it should be considered as a possible approximation to a more general
theory, where the gravitational interaction of general relativity and the interactions of the standard model of
particle physics are incorporated to a description with respect to one unified symmetry principle. According
to this interpretation the intersection term would be a consequence of this more general description of nature
as a generalization of the usual description according to the known gauge theories at low energies. The
corresponding generalized algebra represents as supersymmetry a way to relate internal and external symmetries
by an algebra extending the special properties of a Lie algebra. The generalization of the interaction of the
gravitational field with Yang-Mills fields according to the intersection of Yang-Mills theory with the gauge
description of general relativity could give an alternative explanation or at least a partial explanation to
astrophysical and cosmological phenomena, which are usually interpreted as consequences of the existence of
dark matter. This seems to be quite plausible, since the modification of the gravitational interaction just
concerns the interaction of gravity with gauge bosons and not with fermionic matter and the interaction of
the gauge bosons of the standard model with gravity is not explored very well empirically. Of course, the
generalization can also be applied to the simplest case of electromagnetism with $U(1)$ gauge group.
In this case the interaction of gravity with electromagnetic radiation would be modified. This could
become interesting in particular with respect to cosmology. Accordingly under certain conditions the
presented theory could even yield a possible explanation for the acceleration of the expansion rate
of the universe.
\newpage
%$Acknowledgement$:\\ 

\end{document}